\documentclass[12pt]{article}
\usepackage[cp1251]{inputenc}
\usepackage[english]{babel}
\usepackage{graphicx,graphics}
 0
\begin{document}
\title{Covariant forms of Lax one-field operators: from Abelian to non-commutative}
\author{  S.B. Leble
\small \\ Theoretical Physics and Mathematical Methods Department,
\small\\ Gda$\acute{n}$sk University of Technology, ul,
Narutowicza 11/12, Gda$\acute{n}$sk, Poland,
\small \\ leble@mifgate.pg.gda.pl \\  \\[2ex] }
\maketitle

\renewcommand{\abstractname}{\small Abstract}
 \begin{abstract}
 Polynomials in differentiation
  operators are considered. The
Darboux transformations covariance determines non-Abelian entries
to form the coefficients of the polynomials. Joint covariance of a
pair of such polynomials (Lax pair) as a function of one-field is
studied. Methodically, the transforms of the coefficients are
equalized to Frechet derivatives (first term of the Taylor series
on prolonged space) to establish the operator forms.  In the
commutative (Abelian) case that results in binary Bell (Faa de
Bruno) differential polynomials having natural bilinear
representation. The example of generalized Boussinesq equation is
studied, the chain equations for the case are derived. A set of
integrable non-commutative potentials and hence nonlinear
equations is constructed altogether with explicit dressing
formulas.
\end{abstract}

\thispagestyle{empty}

\section {Introduction.}
Investigations of general DT theory in the case of differential operators
\begin{equation}\label{1}
L = \sum_{k=0}^n a_k\partial^k
\end{equation}
with non-commutative coefficients was launched by papers of
Matveev \cite{M}. The proof of a general covariance of the
equation
\begin{equation}\label{EQ}
 \psi_t = L\psi
\end{equation}
with respect to the classic Darboux transformation
\begin{equation}\label{DT}
\psi[1]=\psi'-\sigma \psi
\end{equation}
incorporates the auxiliary relation
\begin{equation}\label{Miu}
\begin{array}{c}
\sigma_t = \partial r  + [r,\sigma], \\
 r=\sum_0^Na_nB_n(\sigma),
\end{array}
\end{equation}
where $B_n$ are differential Bell (Faa de Bruno \cite{Faa})
polynomials \cite{LeZa}. The relation (\ref{Miu}) generalize
so-called Miura map and became the identity when $\sigma =
\phi'\phi^{-1}$, $\phi$ is a solution of the equation (\ref{EQ}).

Such combinations may be used in the Lax representation
constructions  for  nonlinear  problems. It opens the way to
produce wide classes of solutions of the nonlinear problem. This
known approach \cite{MR93d:35136} intensely develops nowadays.
Examples of non-Abelian and non-local equations, integrable by DT
was considered in \cite{S}, \cite{LeSa}. Some of them were
reviewed and developed in the book \cite{MR93d:35136}. The general
approach was developed for difference operator (see again the
papers cited above) and recently generalized for wide class of
polynomials of automorphism on a differential ring
\cite{MR99c:58144}.

The choice of the jointly covariant combinations introduces addition problems of the
appropriate choice of potentials on which the polynomial coefficients depend \cite{
MR93h:58142}. This problem was recently discussed in
 \cite{Le}, where a method of the conditions account was developed.
 Covariant combinations of (generalized) derivatives and potentials
may be hence classified for linear problems. The task is
intimately connected with other reduction problems of Lax operator
representation of nonlinear equations. In two words, having the
general statement about covariant form of a linear polynomial
differential operator that determine transformation formulas for
coefficients (Darboux theorem and its Matveev's generalizations),
the consistency between two such formulas need the special
definition of potentials. For example, the second order scalar
differential operator has the only place for a potential and the
covariance generate the classic Darboux transformation for it. The
place that this potential take in the second  equation of the KdV
Lax pair needs special investigation.

In scalar case such one-potential construction have been studied
in \cite{LLS} and developed for higher KdV and KP equations
\cite{LLoS}. It was found that the result is conveniently written
via such combinations of differentiation operator and exponential
functions of the potential as  Binary Bell Polynomials (BBP)
\cite{MR96m:58110}. The principle and result is reproduced and
developed in the Sec. 1.2 of this paper to give more explanation.

The whole construction in general (nonAbelian) case
 could be similar, a bit more complicated, but
much more rich and promising. The theory could contain two
ingredients.

i)The first one would be non-Abelian Hirota construction in the
terms of the mentioned binary Bell polynomials \cite{MR96m:58110}.
On the level of general expressions and applications some obstacle
appears, e.g. an extension of addition formulas
\cite{MR96m:58110}) to the nonAbelian case.

ii) The second way relates to some generalized polynomials that
could be produced as covariant combinations of  operators with a
faith that observations from Abelian theory could be generalized.
Namely the case we would discuss in this paper.

Even for the minimal (first order in the D-operator) examples one
arrives to the operator ZS problems that contain many interesting
integrable models. It is seen already from the point of view of
symmetry classification \cite{MR1776509}. So, the link to DT
covariance approach allows to hope for a realization of the main
purpose - construction of covariant polynomials and their
enumeration and use in the soliton equations theory.

We would begin from the example, using notations from quantum
mechanics to emphasize the non-Abelian nature of the
consideration.
 The operators $\rho$ and $H$ could play the roles of density
matrices and Hamiltonians, respectively, but one also can think of
them as just some operators without any particular quantum
mechanical connotations. The approach establishes the covariance
with respect to bDT \cite{MR99c:35235} of rather general Lax
system for the equation
 $$
 -i\rho_t=[H,h(\rho)],
 $$
where $h(\rho)$ - analytic  function, in some sense -"Abelian",
the function to be defined by Taylor series \cite{LCUK}.
  More exactly it is shown that the following statement takes place

  \medskip\noindent

{\bf Theorem.} {\it Assume $\langle\psi|$ and $|\varphi\rangle$
are solutions of the following (direct) equations
$$
z_\nu\langle\chi|
 =
\langle\chi|(\rho -\nu H),\label{2-a}\\
$$
$$
-i\langle\chi_t| =
\frac{1}{\nu}\langle\chi| h(\rho) ,\label{2-b}\\
$$
and $|\varphi\rangle$ $|\varphi\rangle$  is for conjugate pair
{\it Here $\rho$, $H$ are operators left-acting on a ``bra"
vectors $\langle\psi|$ associated with an element of a Hilbert
space and transforms (bDT) complex numbers $\lambda$, $z_\lambda$
are independent of $t$.
 and $\langle\psi_1|$,
$\rho_1$, $h(\rho)_1$ are defined by
\begin{equation}
 \langle\psi_1|
 =
\langle\psi|\Big(1+\frac{\nu-\mu}{\mu-\lambda}P\Big),\label{7.5.DT}
\end{equation}
\begin{equation}
\rho_1
 =
T \rho T^{-1},\label{7.5.DTU}\\
h_1(\rho) = T h(\rho)T^{-1}, \qquad T = \Big(
1+\frac{\mu-\nu}{\nu}P\Big),
\end{equation}
where
  $P =  |\varphi \rangle \langle \chi|\varphi \rangle \langle\chi|$.
Then the pairs
  are covariant:}}
$$
z_\lambda\langle\psi_1| = \langle\psi_1|(\rho_1 -\lambda H),
\qquad -i\langle\dot\psi_1| =
\frac{1}{\lambda}\langle\psi_1|h_1(\rho).
$$

The Lax-pair representation and Darboux covariance properties of
this equation have been established in \cite{LCUK}. The cases
$f(\rho)=i\rho^3$ and $f(\rho)=i\rho^{-1}$ were considered in
\cite{MR1776509}. A step to further generalizations for
essentially non-Abelian functions, e.g.  $h(X) = XA+AX, [A,X] \neq
0$, is studied in \cite{MR2000c:82052}. The case is the
development of the matrix representation of the Euler top  model
\cite{MR56:13272}.

This example of the theory  is more close to the spirit
of the sec.
 2.4 and more achievements are demonstrated in \cite{CU},
 where abundant set of integrable equations is listed. The list is in a
partial correspondence with
 \cite{MR1776509},
 and give the usual for the DT technique link to solutions via the
iteration procedures or dressing chains. The results show how the
"true" non-Abelian functions appear in the context of the
covariance conditions application. On the way of this
specification we use the notion that is similar to one for
automorphic function.

\section{Covariance principle equations}

\subsection{One-field Lax pair for Abelian case. Covariance equations}

 First we would reproduce the "Abelian" scheme, generalizing the study of the example of the
 Boussinesq equation \cite{Le}. To start
 with the search we should fix the number of fields.
 Let us consider the third order operator (\ref{1}) with coefficients $b_k, k=0,1,2,3$,
 reserving $a_k$ for the second operator in a Lax pair.  Suppose, both
 operators depend on the only potential function $w$.
We would restrict ourselves by the case of $b_{ix} = 0$ and the
special choice $ b_3 = 1$ , $b_1 = b(w)$, $b_0 = G(w)$.  The
problem we consider now may be formulated as follows: To find
restrictions on the coefficients $b_2(t), b(w,t), G(w)$ compatible
with DT transformations rules of the potential function $w$
induced by DT for $b_i$.
 The standard DT (\ref{DT})
for the third order operator coefficients (Matveev generalization
\cite{M}) yields (we denote $\partial f=f'$)
\begin{equation}\label{dt1}
b_2[1] = b_2 + b_3',
\end{equation}

\begin{equation}\label{dt2}
b_1[1] = b_1 + b_2' + 3b_3 \sigma',
\end{equation}

\begin{equation}\label{dt3}
b_0[1] = b_0 + b_1' +  \sigma  b_2' + 3b_3 (\sigma \sigma' +
 \sigma''),
\end{equation}
having in mind that the "elder" coefficient $b_3$  does not
transform. Note also, that $b_3'=0$ yields invariance of the
coefficient $b_2$.

 The general idea of DT form-invariance may be
realized considering the coefficients transforms to be consistent
with respect to the fixed transform of $w$. Generalizing the
analysis of the third order operator transformation \cite{Le}, one
arrives at the equations for the functions $b_2(t),b(w,t),G(w)$.
The covariance of the spectral equation
\begin{equation}\label{sp}
b_3 \psi_{xxx} + b_2(t)\psi_{xx} + b(w,t) \psi_{x} + G(w) \psi  =
\lambda \psi
\end{equation}
may be considered separately, that leads to the link between $b_i$
only. We study the problem of the (\ref{sp}) in the context of Lax
representation for some nonlinear equation, hence the  covariance
of the second Lax equation is taken into account from the very
beginning. We name such principle as the "{\it principle of joint
covariance}" \cite{MR93h:58142}. The second (evolution) equation
of the case is :
\begin{equation}\label{ev}
\psi_t = a_2(t)\psi_{xx} + a_1(t)\psi_x +  w \psi,
\end{equation}
 with the operator in the r.h.s. having again the form of
 (\ref{1}).  We do not consider here a dependence of $a_i, b_i$  on
 $x$ for the sake of brevity, leaving this interesting question to the next paper.

   If one consider the $L$ and $A$ operators of the form (1), specified in equations
    (\ref{sp}) and (\ref{ev}).
as the Lax pair equations, the DT of $w$ implied by the covariance
of (\ref{ev}),
 should be compatible with DT formulas of
  both   coefficients of (\ref{sp}) depending on the only  variable
  w.
$$
a_2[1] = a_2 = a(x,t),
$$
$$
a_1[1] = a_1(x,t) + Da(x,t)
$$
\begin{equation}\label{DTa}
a_0[1] = w[1] = w +  a_{1}' + 2 a_2  \sigma' + \sigma a_2'
\end{equation}

Next important relations being in fact the identities in the
  DT transformation theory \cite{LeZa}, see the introduction, are the particular cases
  of the generalized Miura map,(\ref{Miu}):
 \begin{equation}\label{M1}
 \sigma_t = [a_2( \sigma^2 + \sigma_x) +a_1\sigma + w]_x
\end{equation}
 for the problem (\ref{ev}) and, for the (\ref{sp})
\begin{equation}\label{M2}
\sigma^3 + 3\sigma_x\sigma + \sigma_{xx} + b(w,t)\sigma + G(w) =
const;
\end{equation}
$\phi$ is a solution of both Lax equations.
 Suppose now that the coefficients
 of the operators are analytical functions of w together with
 its derivatives (or integrals) with respect to x (such functions are named functions
 on prolonged space \cite{O}).
 For the coefficient $G$ it means
 \begin{equation}\label{pro}
 G = G(\partial^{-1}w, w, w_x,...\partial^{-1}w_t,w_t,w_{tx},...),
\end{equation}
The covariance condition is obtained for the Frech$\hat{e}$t
derivative (FD) of the function $G$ on the prolonged space, or the
first terms of
 Taylor series for (\ref{pro}), read
\begin{equation}\label{FD}
 G(w +  a_{1}' + 2 a_2  \sigma' + \sigma a_2') = G(w) +  G_{w_x}(a_1'+ 2a_2\sigma'+
 \sigma a_2')'
  + G_{\partial^{-1}w_t}( a_{1t}+2a_2\partial^{-1}(\sigma_t')+\partial^{-1}(\sigma)+ ... ,
\end{equation}
 where we show only terms of further importance; the expression simplifies
if $a_2$ does not depend on $x$. Quite similar condition one have
for
 the$b_1 = b(w,t)$, with which we would start.
 In the analogy with the expressions (\ref{FD},\ref{dt2}) one obtains
\begin{equation}\label{CC1}
 b_2' + 3b_3 \sigma' =  b_w(a_{1}' + 2 a_2  \sigma' + \sigma
a_2') + b_{w'}(a_{1}' + 2 a_2  \sigma' + \sigma a_2')' ... .
\end{equation}
This equation we name  the (first) {\it "joint covariance
equation"} that guarantee the consistency between transformations
of the coefficients of the Lax pair (\ref{ev}), (\ref{sp}).
 In the frame of our choice $a_2'=0$, the equation simplifies
\begin{equation}\label{CC11}
\begin{array}{c}
 3b_3  = 2b_w  a_2, \\
 b_2' = b_w a_{1}',
\end{array}
\end{equation}
or
\begin{equation}\label{CC11'}
\begin{array}{c}
  b_w = 3b_3 /2a_2 \\
 b_2' =   a_{1}'3b_3 /2a_2.
\end{array}
\end{equation}
So, if one wants to save the form of
 the standard DT for the variable $w$
 (potential) the simple comparison of both transformation formulas gives for
 b(w) the following connection (with arbitrary function)
\begin{equation}\label{b}
  b(w,t)  = 3w/2 + \alpha(t).
\end{equation}

 Equalizing the expansion (\ref{FD}) with the
 transform of the $b_0=G(w)$ yields:
\begin{equation}\label{CC2}
 b_1' +  \sigma  b_2' + 3b_3 (\sigma^2/2  +
 \sigma')' = G_{w_x}(a_1'+ 2a_2\sigma'+\sigma a_2')'
  + G_{\partial^{-1}w_t}[
  a_{1t}+2\partial^{-1}(a_2 \sigma_t')+ \partial^{-1}(\sigma a_2')_t]+...
\end{equation}
 This second {\it
 "joint covariance equation"} also simplifies when $a_2' = 0$:
\begin{equation}\label{CC2'}
3b_3w'/2a_2 + 3b_3 (\partial^{-1}\sigma_t-w )'/a_2 +
3b_3\sigma''/2 = G_{w_x}( 2a_2\sigma')' + G_{\partial^{-1}w_t}[
2a_2 \sigma_t]+...
\end{equation}
 when (\ref{b}) is accounted. Note, that the "Miura" (\ref{ev}) linearizes
 the FD with respect to $\sigma$. Finally,
\begin{equation}\label{G}
 \begin{array}{c}
     G_{w_x} = 3b_3/2a_2\\
  G_{\partial^{-1}w_t} = 3b_3/2a_2^2  \
 \end{array}
\end{equation}

 Compare with the formula  $w[1] = w + 2\sigma_x$.
The transformation for the potential w
  follows from the last equation of the
 system (\ref{dt3}), i.e.,
\begin{equation} \label{3.31}
 G[1] = G + 3w_x/2 +  3(\sigma^2/2 +
 \sigma_{x})_x.
\end{equation}
 Such  equation determine the functional dependence of G(u) and
  we would name such equations as
 {\it joint covariance equations}. We
  see that further analysis is necessary due to such constraint
(reduction)
 existence.
 Then, similar to the case of KdV equation,
 the covariant equations (\ref{sp},\ref{ev})
\begin{equation}\label{a1}
 a_{1t} + a_2a_1''+a_1a_1' = 0,
\end{equation}
which get the form of the Burgers equation after (\ref{CC11'})
account.  Finally the "lower" coefficient of the third order
operator is expressed by
\begin{equation}\label{G}
G(w,t) = 3b_3w_x/2a_2 +  3b_3 a_1'\partial^{-1}w /2(a_2)^2 +
3b_3\partial^{-1}w_t /2a_2^2.
\end{equation}
{\bf Statement 1} The expressions (\ref{ev}, \ref{sp}, \ref{b},
\ref{G}) define the covariant Lax pair when the constraints
(\ref{a1}, \ref{CC11'}) are valid.

 Such equation for (\ref{ev})
  compare with
 Riccati  equation (stationary version)
 for the second order spectral problem
 corresponding to KdV provided by the name of Miura equation. If one would use the
 equation (\ref{3.32}) in (\ref{3.31}), the time-derivative of w
 appear. Moreover, the further analysis shows that the case we study need to widen the
 functional dependence in u, namely we should include not only derivatives of w with
 respect to x,
 but integrals (inverse derivatives) as arguments of the potential.
 Let us  introduce
 analytical function G denoting the coefficient $b_0$.

The DT transform of $G$ after substitution
 of (\ref{3.31}) gives
\begin{equation}\label{3.36}
 G(w) + 3w_x/2 +  3(- \sigma_t - w_x)/2 +
 3\sigma_{xx}/2.
\end{equation}
 Equalizing (\ref{3.36}) and (\ref{3.31}) yields
 $$
 G_{w_x} = 3/4; G_{\partial^{-1}w_t} = - 3/4.
 $$
 That leads to the exact form of the Lax pair for the Boussinesq equation from \cite{MR93d:35136} for the choice
 of $\alpha = -3/4$.

{\bf Remark 1.} We cut the Frech$\hat{e}$t differential formulas
on the   level that is  necessary  for the minimal flows. The
account of higher terms leads to the whole hierarchy
 \cite{LSoS}.

 {\bf Remark 2.} We cut the Frech$\hat{e}$t differential formulas on the
  level that is
 necessary  for the minimal flows. The account of higher terms lead to higher
 flows (higher KdV, for example) \cite{LSW}.

\section{The solitons of the generalized Boussinesq and
dressing chain for the Boussinesq equation}

To produce a simplest soliton solution of the generalized
Boussinesq equation, it is enough to start from zero potential in
the Lax pair equations (\ref{sp},\ref{ev}).
\begin{equation}\label{LA0}
\begin{array}{c}
  b_3 \psi_{xxx} + b_2(t)\psi_{xx} + \alpha(t) \psi_{x}  =
\lambda \psi \\
  \psi_t = a_2(t)\psi_{xx} + a_1(t)\psi_x
\end{array}
\end{equation}

The seed solution should satisfy both Lax equations (\ref{LA0})The
dressing formula for the zero seed  potential (\ref{DT}) is
standard and includes this only function.
\begin{equation}\label{sol1}
  w_s = a_{1}' + 2 a_2  \sigma' + \sigma a_2' = a_{1}'+ 2a_2\log_{xx}\phi(x,t)
\end{equation}

 Going to the dressing chain, we use the method from \cite{Le}.
 We would restrict ourselves to
 the case of constant $b_2 = 0$, $ b_3 = 1$ ,  $b_1 = b$, $b_0 = u$. The
 general construction is quite similar.

 the covariant equations (\ref{ev},\ref{sp}) are accompanied by the
 following
 equation
\begin{equation}\label{3.32}
  \sigma_t = -( \sigma^2 + \sigma_x)_x - w_{x}
\end{equation}
 for the problem (\ref{ev}) and
\begin{equation}\label{3.33}
  \sigma^3 + 3\sigma_x\sigma + \sigma_{xx} + b\sigma + G = const,
\end{equation}
 for (\ref{sp}), see (\ref{M1}), compare with (\ref{M2}) that was  Riccati  equation (stationary version)
 for the second order spectral problem
 corresponding to KdV. If one would use the equation (\ref{3.32}) in (\ref{3.31}), the time-derivative of w
 appear.

Namely the "Miura" equations (\ref{3.32}, \ref{3.33} together with
the DT formula \ref{7.5.DT}
\begin{equation}\label{CH1}
  w_{n+1} = w_n + 2ln_x \sigma_n
\end{equation}
form the basis to produce the DT dressing chain equations.

We express the iterated  potential $w_n$ from \ref{3.32}
\begin{equation}\label{CH2}
 - \sigma_{nt} -( \sigma_n^2 + \sigma_{nx})_x = w_{nx}
\end{equation}

and substitute it into {\ref{CH1}}. The first dressing chain
equation
\begin{equation}\label{C}
\sigma_{n+1,t} - \sigma_{nt}  =  ( \sigma_{n+1}^2 +
\sigma_{n+1}')' - ( \sigma_n^2 - \sigma_{n}')'.
\end{equation}
Next chain equation is obtained when one plugs the potential from
(\ref{CH2}) to the iterated (\ref{M2s})
\begin{equation}\label{C2}
\sigma_n^3 + 3\sigma_n'\sigma_n + \sigma_n'' + (-3u_n/2 +
\alpha)\sigma_n + - 3u_n'/4+3\partial^{-1}u_{nt} = c_n.
\end{equation}

 \section{Non-Abelian case. Zakharov-Shabat (ZS) problem.}

\subsection{Compatibility condition.}

In the case $a_2' = 0 $ by which we have restricted ourselves, the
Lax system (\ref{ev},\ref{sp}) produces the following
compatibility conditions:
\begin{equation}\label{CoC}
\begin{array}{c}
2a_2b_3' = 3b_3a_2', \\
  b_{3t} = a_2b_3'' + 2a_2b_2' + a_1b_3' - 3b_3a_2'' - 3b_3a_1' - 2b_2a_2' \\
  b_{2t} =   a_2b_2'' +2a_2b_1' +a_1b_2' - b_3a_2''' - b_2a_2'' - b_1a_2' - 3b_3a_1'' - 2b_2a_1' - 3b_3a_0' \\
  b_{1t} =  a_2b_1'' + a_1b_1' - b_3a_1''' - b_2a_1'' - b_1a_1' - 3b_3a_0'' - 2b_2a_0' + 2a_2b_0'\\
    b_{0t} = a_1b_0' - b_1a_0' + a_2b_0'' - b_2a_0'' - b_3a_0'''
\end{array}
\end{equation}
In the particular case of $a_2'  = 0$  we have at once the direct
corollary of the first of equalities (\ref{CoC}) $b_3' = 0$; in
the rest of the equations the restriction (\ref{CoC}) is taken
into account.

 \subsection{Compatibility conditions for two general ZS problems}
Let us list equations that appear as compatibility condition of two first order
operators with coefficients to form a nonabelian set. We recall these equation just to examine the complete set of them from the DT invariance point of view.
The pair we study is
\begin{equation}\label{ZS1}
  \psi_t = (a_0+a_1 D)\psi,
\end{equation}
\begin{equation}\label{ZS2}
  \psi_y = (b_0+b_1 D)\psi,
\end{equation}
where $a_i, b_i$ are functions of $x,y,t$ and let D be a
differentiation with respect to $x$ operator.
 The equations of compatibility are
 \begin{equation}\label{ce1}
  [b_1,a_1=0]
\end{equation}

\begin{equation}\label{ce2}
 a_{1y} - b_{1t} + [a_0,b_1]+[a_1,b_0]+a_1b_1' - b_1a_1'=0.
\end{equation}

\begin{equation}\label{ce3}
 a_{0y} - b_{0t} + [a_0,b_0]+a_1b_0'- b_1a_0'=0.
\end{equation}
The "'"  denotes the derivative with respect to x.

If the coefficients $a_1, b_1$ do not depend on $x,y,t$, the first
two equations simplify.
\begin{equation}\label{ces1}
  [b_1,a_1=0]
\end{equation}

\begin{equation}\label{ces2}
  [a_0,b_1]+[a_1,b_0] =0.
  \end{equation}
Analyzing the conditions, one sees, that the direct proof of the
heredity of (\ref{ce2}) is a corollary of Jacoby identity.

If one changes the differentiation operator $D$ in (\ref{ZS1,ZS2}) to the shift operator $T$,
 the equations
are changing as follows

In the next generalization \cite{MR99c:58144}, the operator  $T$
can be
 considered as automorphism in a differential ring.
Some direct generalization of (\ref{ce1,ce2,ce3}) is achieved, if
the operator $D$ is considered as abstract differentiation. See.
for example, \cite{MR93h:58142} where $D$ is a commutator.

\subsection{Covariance equations}
Let us change notations

\begin{equation}\label{ZS11}
  \psi_t = (J+uD)\psi,
\end{equation}
where the operator J does not depend on $x,y,t$ and the potential
$a_0 = u=u(x,y,t)$ is a function of all variables. The operator D now is a differentiation by x. The
transformed potential
 \begin{equation}\label{DT}
 \tilde{u} = u + [J,\sigma],
\end{equation}
the $\sigma=\phi_x\phi^{-1}$ is defined by the same formula as
before. The covariance of the operator in (\ref{ZS1}) follows from
general transformations of the coefficients of a polynomial
\cite{MR93d:35136}. The operator J does not transform.

Suppose the second operator of a Lax pair has the same form, but
with different entries and derivatives.
\begin{equation}\label{ZS12}
  \psi_y = (Y+wD)\psi,
\end{equation}
where the potential $w=F(u)$ is a function of the potential of the
first (\ref{ZS11}) equation. The principle of joint covariance
\cite{MR93h:58142} hence reads:
  \begin{equation}\label{DT2}
 \tilde{w} = w+[Y,\sigma] = F(u+[J,\sigma]),
\end{equation}
with the direct corollary
 \begin{equation}\label{CP}
F(u)+[Y,\sigma] = F(u+[J,\sigma]).
\end{equation}
 It implies the same functional
dependence of the coefficient $F$ on $u$ before and after
transformation. So, the equation (\ref{CP}) defines the function
F(u), we shall name this equation as {\bf joint covariance equation}.
In the  case of abelian algebra we used the Taylor
series (generalized by use of a Frechet derivative) to determine
the function. Now some generalization is necessary. Let us make some general remarks.

A class operator-valued function  F(u) of an operator u in Banach space
may be considered as a generalized Taylor series with coefficients
that are expressed in terms of Frechet derivatives \cite{Enc}. In
a sense of the space norm the linear in u part of the series approximate the
function
$$
F(u) = F(0) + F'(0)u + ...
$$
The representation is not unique and the similar expression
$$
F(u)= F(0)+ u \hat{F}'(0)  + ...
$$
may be introduced (some fundamentals about the definitions are given in Appendix).  Both expressions  however are not Hermitian, hence not siutable of
a majority of physical models. It means, that the class or such operators is too restrictive. To explain what we
have in mind, let us
consider examples.

\subsection{Important example}

From a point of view of the  physical modelling the following
approximation
$$
 F(u) = H^+u+uH
$$
is preferable for the important class of  Hermitian operators. The
case of the Hermitian H is included. Such models could be applied
to quantum theories: introduction of this approximation is similar
to "phi in quadro" (Landau-Ginzburg) model \cite{}. So, let it be
 \begin{equation}\label{LG}
 F(u)=Hu+uH,
\end{equation}
by a direct calculation in (\ref{CP}) one arrives at the equality
 \begin{equation}\label{CP1}
 [Y,\sigma] = H[J,\sigma]+[J,\sigma]H.
\end{equation}
The obvious choice for arbitrary $\sigma$ is $Y=H^2, J=H$.

The compatibility conditions (\ref{ce2,ce3} for (\ref{ZS11}) and
(\ref{ZS12}), when $\sigma$ is $Y=H^2, J=H$ yields
\begin{equation}\label{KK}
 \begin{array}{c}
 [J,J^2]=0 \\
\left[J,u\right]H + \{u[J,H]\}+[u,H]J = 0 \\
    u_y-Hu_t-u_tH + [u,H]u+u[u,H] + JHu_x+Ju_xH+HJu_x = 0 \
 \end{array}
\end{equation}
So this important case of [H,J]=0 gives
\begin{equation}
  \begin{array}{c}
     [J,u]H+[u,H]J =0 \\\label{EQ}
    u_y-Hu_t-u_tH + [u,H]u+u[u,H] + JHu_x+Ju_xH+HJu_x = 0 \
  \end{array}
\end{equation}
Both condition are covariant, the first of them is a DT-hereditary
constraint, that may be checked directly. We hence obtain a class
of DT-integrable equations/Lax pairs.
\begin{equation}\label{GMan}
   u_y-\{H.u\}_t + [u^2,H] + JHu_x+Ju_xH+HJu_x = 0,
\end{equation}
 It is seen that in the case J=H, the first of equations (\ref{EQ}) is valid
 identically.

 if the potential does not
 depend on t it is reduced to the next equation:
 \begin{equation}\label{GMan2}
   u_y  + [u^2,J]  + J^2u_x+Ju_xJ+J^2u_x = 0,
\end{equation}
and x-independence yields the generalized Euler top equations

\begin{equation}\label{Euler}
   u_y  + [u^2,J] = 0,
\end{equation}
which  Lax pair  (\ref{ZS11},\ref{ZS12}) with $Y=J^2$ was found by
Manakov \cite{MR56:13272}. So, we used the compatibility condition
to find the form of integrable equation and reduction
 tracing the simplifications appearing for the subclasses of covariant potentials. While doing
 this we also check the invariance of the equation and  heredity of te constraints.

\subsection{Covariant  combinations of symmetric polynomials}
The next natural example appears if one examine the link
(\ref{CP1}).
$$
P_2(H,u)=H^2u+HuH+uH^2
$$
The direct substitution in the covariance and compatibility
equations  leads to covariant constraint that goes to identity if,
$Y=H^3,  J=H$.

It is easy to check more general $Y=J^n, J=H$ connection
possibility that leads to the covariance of
$$
P_n(H,u)=\sum_{p=0}^nH^{{n-p}}uH^{p}.
$$
Such observation was exhibited in \cite{LC}. On the way of a
further generalization let us consider
\begin{equation}\label{comb}
f(H,u)=Hu+uH + S^2u+SuS+uS^2
\end{equation}
Plugging (\ref{comb}) as $F(u) = f(H,u) $ into (\ref{CP}),
representing $Y = AB + CDE$ yields
$$
A[B ,\sigma]+ [A ,\sigma]B+CD[E ,\sigma]+C[D ,\sigma]E+[C
,\sigma]DE=H[J ,\sigma]+[J ,\sigma]H + S^2 [ J,\sigma]+ S[J
,\sigma]S+ [J ,\sigma]S^2.
$$
The last expression turns to identity if $A=B=J=H, C=\alpha H, D =
\alpha H, D = \alpha H, S = \beta H$  and $ [\alpha,
H]=0,[\beta,H]=0$ with the link $\alpha^3=\beta^2$.

   {\bf Statement} {\it Darboux covariance define a class of homogeneous
   polynomials $P_{n}(H,u)$, symmetric with respect to cyclic permutations. A linear
   combination of such polynomials $\sum_{n=1}^{N}\beta_nP_n(H,u)$ with the
    coefficients commuting with $u,H$ is also
   covariant, if the element $Y=\sum_{n=1}^{N}\alpha_n H^(n+1)$ and $\alpha_1=\beta_1=
   1, \alpha_n^{n+2}=\beta^{n+1}, n \neq 1 $.}
A proof could be made by induction that is based on homogeneity of
the $P_n$ and  linearity of the constraints with respect to u. The
functions $F_H(u) = \sum_0^{\infty} a_n P_{Hn}(u)$ satisfy the
constraints if the series converges.

\section{Conclusion}
The equation () generalize Boussinesq equation and has direct
physical sense for both case of waves launched by initial or
boundary problem. In the first case t-dependence of coefficients
means the external conditions varied with time. In the second case
the T and x coordinates interchange: the coefficients dependence
may be interpreted as conditions that are changed when a wave
propagates (e.g. a bottom slope for the surface waves). The main
result of this paper is the covariant equation (\ref{CP}) or its
version (\ref{CP1}). A class of potentials from \cite{CU} contain
polynomials $P_{Hn}(u)$ and give alternative expressions
 for it. The linear combinations, introduced here could better
 reproduce physical situation of interest.

 The work is also supported by  KBN grant 5P03B 040 20.
  
\section{Appendix. Right and left Frech$\hat{e}$t derivatives}

The notion of a derivative an operator by other one is defined in
a Banach space $B$
\cite{G}. Two specific features  in the case of a
operator-function $F(u)\in\ B\,u\in B$  should be taken into
account: a norm choice when a limiting procedure is made and the
nonabelian character of expressions while the differential and
difference introduced.

{\bf Definition. } Let a Banach space B have a structure of a
differential ring. Let F be the operator from B to B' defined on
the open set of B. The operator is named the left-differentiable
in $u _0 \in B$ if there exist a linear restricted operator
$L(u_0)$, acting also from B to B' with the property
\begin{equation}
 L(u_0+h)-L(u_0) = L(u_0)h +\alpha(u_0,h),\\
 ||h||\rightarrow 0,
\end{equation}
  where
$||\alpha(u_0,h)||/||h||\rightarrow 0.$ The operator
$L(u_0)=F'(u_0)$ is referred as the operator of the (strong) left
derivative of the function $F(u)$. The right derivative
$\hat{F}'(u_0)$ could be defined by the similar expression  and
conditions, if one changes $Lh \rightarrow h\hat{L}$ in the
(\ref{A}).

The Gatoux derivative

\end{document}